\documentclass[useAMS,usenatbib]{mn2e}  
\usepackage{amssymb,amsmath}
\usepackage{amsfonts}
\usepackage{graphicx}
\usepackage{booktabs}
\usepackage{tabularx}

\numberwithin{equation}{section}

\def\d{{\rm d}}

\newcommand{\p}{\partial}

\newcommand{\vx}{{\mathbf x}}
\newcommand{\vJ}{{\mathbf J}}
\newcommand{\vA}{{\mathbf A}}

\renewcommand{\[}{\left[}

\begin{document}

\title[Bar formation in cuspy models]{On the bar formation mechanism \\in galaxies with cuspy bulges}
\author[E. V. Polyachenko, P. Berczik, A. Just]{E.~V.~Polyachenko, $^1 $\thanks{E-mail: epolyach@inasan.ru}\ \ \ P.~Berczik, $^{2,3,4} $\thanks{E-mail: berczik@mao.kiev.ua}\ \ \ A.~Just, $^2 $\thanks{E-mail: just@ari.uni-heidelberg.de}\\
     $^1 ${\it\small Institute of Astronomy, Russian Academy of Sciences,}
          {\it\small 48 Pyatnitskya St., Moscow 119017, Russia}\\
     $^2 ${\it\small Astronomisches Rechen-Institut am ZAH, }
          {\it\small M\"onchhofstr. 12-14, D-69120 Heidelberg, Germany} \\
     $^3 ${\it\small Main Astronomical Observatory, National Academy of Sciences of Ukraine, MAO/NASU,}
          {\it\small 27 Akad. Zabolotnoho St. 03680 Kyiv, Ukraine} \\
     $^4 ${\it\small National Astronomical Observatories of China, Chinese Academy of Sciences, NAOC/CAS}\\
          {\it\small 20A Datun Rd., Chaoyang eDistrict Beijing 100012, China}}



\maketitle

\begin{abstract}

We show by numerical simulations that a purely stellar dynamical
model composed of an exponential disc, a cuspy bulge, and an NFW
halo with parameters relevant to the Milky Way Galaxy is subject to
bar formation. Taking into account the finite disc thickness, the
bar formation can be explained by the usual bar instability, in
spite of the presence of an inner Lindblad resonance, that is
believed to damp any global modes.
The effect of replacing the live halo and bulge by a fixed external
axisymmetric potential (rigid models) is studied. It is shown that
while the e-folding time of bar instability increases significantly
(from 250 to 500 Myr), the bar pattern speed remains almost the
same. For the latter, our average value of 55 km/s/kpc agrees with
the assumption that the Hercules stream in the solar neighbourhood
is an imprint of the bar--disc interaction at the outer Lindblad
resonance of the bar.
Vertical averaging of the radial force in the central disc region
comparable to the characteristic scale length allows us to reproduce
the bar pattern speed and the growth rate of the rigid models, using
normal mode analysis of linear perturbation theory in a razor thin
disc. The strong increase of the e-folding time with decreasing disc
mass predicted by the mode analysis suggests that bars in galaxies
similar to the Milky Way have formed only recently.

\end{abstract}


\section{Introduction}

For a long time bars in disc galaxies were explained by a so-called
bar mode instability, discovered in the first $N$-body simulations
of soft-centred stellar models \citep{M70, H71}. However,
applicability of this phenomena to models with cusps, for which a
volume density $\rho \sim r^{-\alpha}$ with positive index $\alpha$
in the centre, has been questioned for two reasons. First, simple
estimates show that in the case of interest the angular velocity
$\Omega(R)$ and epicyclic frequency $\kappa(R)$ are $\propto
r^{-\alpha/2}$, which indicates the presence of the inner Lindblad
resonance (ILR) for any bar pattern speed $\Omega_\textrm{p}$. The
ILR absorbs waves \citep{M71, M74} and damps the formation of global
modes by breaking the feed-back loop for wave amplification
\citep{T81}. Second, Hubble deep field observations show a
significant decline of the fraction of barred galaxies beyond
redshifts $z\sim 0.5$ \citep{A99, M00}. If usual bar formation
begins to operate just after the stellar disc is formed, at these
redshifts we should observe approximately the same fraction of
barred galaxies. According to \citet{S00, K13}, these objections
call for alternative mechanisms of bar formation for cuspy galaxies.

Despite the theoretical input, we know examples of N-body
simulations that incorporate cuspy models which demonstrate an
exponential growth of bars \citep{WPD08}, and here we verify their
conclusion with a larger number of particles. Our main goal is to
explain how such bars are formed in the presence of the ILR.
Throughout this paper we analyse bisymmetric ($m=2$) modes only.

It is still a question whether our Milky Way Galaxy accommodates a
cuspy bulge or not. The rotation curves published earlier
\citep{S09} suggest a bulge with mass about $1.8\cdot 10^{10}$ solar
mass and a weak cuspiness \citep[e.g.,][]{GJ13}. However, more
recent models taking into account non-circular gas motion shift
these estimates toward less massive and more extended bulges
\citep{Ch15}. Anyway, our model galaxy adopts parameters possibly
inherent to the Milky Way. In particular, we assume a weak cusp with
$\alpha\approx 0.5$ in the bulge, and and NFW dark halo
rescaled from the Via Lactea II cosmological simulations \citep{D08,
MJ16}.

N-body simulations of the rigid cuspy models also yield bars.
\citet{A02} noted that bulgeless models with live halos evolve
faster and produce better shaped bars. In terms of unstable modes,
this means larger growth rates. Interesting questions are related to
whether this holds in cuspy models, and if the pattern speeds of
bars change substantially under this replacement. This will be a
by-product of our main investigation.

In stability theory, bar and spiral modes are described by solutions
of the linearised collisionless equations of stellar dynamics
exponentially growing in time \citep{FP84}. At the moment, however,
finding unstable global modes is impossible, if the stellar disc has
a finite thickness, or if particles rather than a fixed external
potential (rigid models) are used to represent the spheroidal
components. Furthermore, except for a few special cases
\citep[e.g.,][]{ER98, T77}, unstable modes were only found reliably
in soft-centred models, i.e. in the absence of a central cusp. A
comparison of the relevant matrix methods for calculation of
unstable global modes can be found in \citet{PJ15}.

The structure of the paper is the following. In Section 2 we
describe the basic model used in Section 3 to show by means of
numerical simulations, with a different number of particles and
calculation schemes, that bar instability really takes place in
cuspy models, both with live and rigid halo and bulge.
Ignoring a difference between thick and razor-thin discs, one can
try to reproduce unstable bar modes of the rigid models by matrix
methods. However, preserving the cuspy velocity curve profile we
obtain no unstable modes. Section 4 considers potential issues and
details of the global mode calculation for stellar discs, and
compares characteristics of the unstable modes for discs with
different vertical scales. In the final part, Section 5, we discuss
the results and outline some perspectives.

\section{The basic model}

Our 3-component model is adopted from \citet{WPD08}, and consists of
the stellar disc, bulge, and dark matter halo. The disc is
exponential, with radial scale $R_\textrm{d} = 2.9$\,kpc, truncation
radius 15 kpc, mass $M_\textrm{d} =4.2\cdot 10^{10}$\,M$_\odot$, and
characteristic height $z_\textrm{d} =300$\,pc ($z_\textrm{d}$ is
defined so that the surface density $\Sigma_d(R) = 2 z_d\rho(R, z=0)
$). The radial velocity dispersion $\tilde \sigma_R$ is exponential,
with central value $\sigma_{R0}=100$\,km/s and radial scale length
$R_{\sigma} = 2R_\textrm{d}$. In the solar neighbourhood
($R=8$\,kpc), the radial velocity dispersion is
$\sigma_{R}=25$\,km/s and the surface density is 50
M$_\odot$/pc$^2$.

The density profile for the bulge is taken in the form
\begin{equation}
\tilde \rho_\textrm{b}(r) = \rho_\textrm{b} \left( \frac r{R_e} \right)^{-p} \textrm{e}^{-b(r/R_e)^{1/n} }\ ,
\label{eq:bulge_dens}
\end{equation}
where $r$ is a spherical radius. Instead of the scale density $\rho_\textrm{b}$, we use the bulge velocity scale
\begin{equation}
\sigma_\textrm{b} \equiv \left\{ 4\pi G n b^{n(p-3)} \Gamma[n(3-p)] R^2_e \rho_\textrm{b} \right\} ^{1/2}\ .
\label{eq:bulge_veld}
\end{equation}
With this definition, $\sigma^2_\textrm{b}$ corresponds to the depth of the gravitational potential associated with the bulge.
For the bulge, free parameters are the S{\'e}rsic index $n=1.11788$, $\sigma_\textrm{b} = 272$ km/s, $R_e = 0.64$ kpc; the derived parameters are $b\simeq 1.92$, $p=1-0.6097/n+0.05563/n^2 \simeq 0.5$, and mass of the bulge $M_\textrm{b}=1.02\cdot 10^{10}$ M$_\odot$.

The target density profile of the halo is a truncated NFW profile
with scale $a_\textrm{h} = 17.25$\,kpc, truncation radius
$r_\textrm{h} = 229.3$ kpc, and total mass
$M_\textrm{h}=1.29\cdot10^{12}$ M$_\odot$. Despite the halo density
distribution being more cuspy than the bulge one, the latter
dominates in the rotation curve down to $R \sim 0.01$ kpc. In
Section 5 we argue that the parameters used in our model give a disc
resolved only up to $R \sim 0.1$ kpc, thus the cusp index for the
model $\alpha = p \simeq 0.5$.

\begin{figure}z
  \centerline{\includegraphics[width = 85mm]{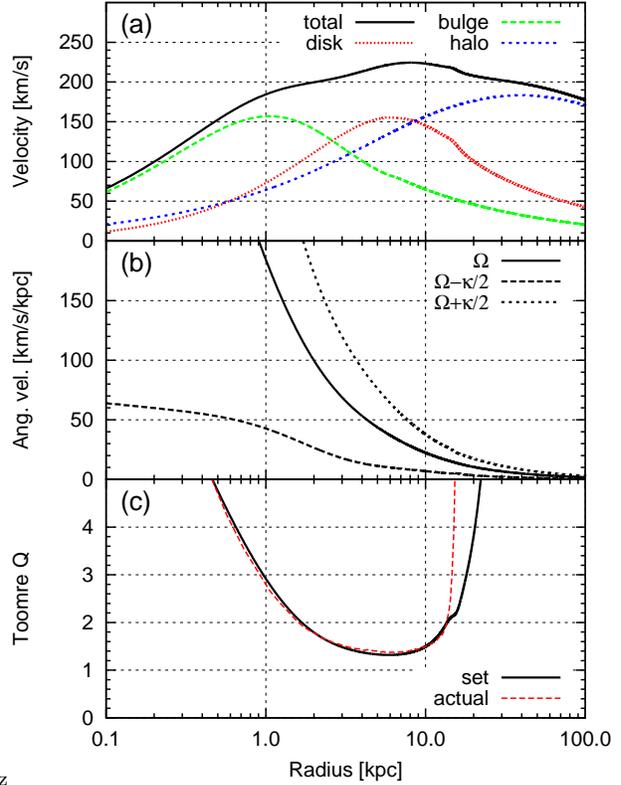}}
  \caption{Initial profiles for the basic model: (a) the total circular velocity and its components due to disc, bulge, and halo; (b) angular velocity $\Omega(R)$ and curves $\Omega(R) \pm \kappa(R)/2$; (c) Toomre $Q$ initially set to GalactICS and one actually obtained in the simulations.}
  \label{fig1}
\end{figure}

The upper panel (a) in Fig.\,\ref{fig1} shows the total circular velocity profile (solid curve) and contributions of separate components. The rotation curve is bulge-dominated at radii $R\lesssim 2.5$ kpc, and halo-dominated at $R > 9$ kpc. At radius $R\approx 6$ kpc, where the disc contribution peaks, the force from the halo is about 2/3 of the force from the disc in the galactic plane.

The angular velocity profile $\Omega(R)$ is presented in panel (b),
along with the curves $\Omega \pm \kappa/2$ which give positions of
the inner and outer Lindblad resonances ILR and OLR respectively.
The ILR for a pattern speed $\Omega_\textrm{p}$ is determined from
the relation
\begin{equation}
\Omega_\textrm{p} = \Omega_\textrm{pr}(R) \equiv \Omega(R) - \frac12 \kappa(R) \ .
\label{eq:res}
\end{equation}
Physically, $\Omega_\textrm{pr}(R)$ determines a precession rate of
nearly circular orbits and plays an important role in the formation
of bars in razor thin discs \citep{EP04}; thus one can refer to it
as a `precession' curve. For the basic model, the curve diverges
weakly at $R \to 0$ as $R^{-\alpha/2}$ with $\alpha \approx 0.5$.

Panel (c) of Fig.\,\ref{fig1} shows the Toomre $Q$ profile
\begin{equation}
Q = \frac{\kappa \tilde \sigma_R}{3.36 G\Sigma_\textrm{d}}\ ,
\label{eq:Qs}
\end{equation}
where $\Sigma_\textrm{d}$ is the disc surface density. The profile remains below 3 in the region $1 < R < 18$ kpc. The minimum $Q_\textrm{min} = 1.32$ is attained at $R= 5.9$ kpc.

\section{N-body simulations}

Initial conditions of stars in single mass simulations have been
generated by the `GalactICS' code provided by \citet{WPD08}.
Numerical calculations have been performed with three different
codes.

A particle--mesh code called {\tt Superbox-10}  (Bien et al. 2013)
solves the Poisson equation by fast Fourier transform. The number of
grid points $ N_g $ for each coordinate is the same and is taken so
that the number of grid cells is comparable with the number of
particles. The code uses grids with three sizes. The largest grid
allows one to simulate an extensive halo and interaction between
galaxies. Medium-size grids are designed to study separate galaxies
and is used for the disc in our case, and the smallest grids are
used to resolve fine structures in galactic centres. In all our runs
$ N_g = 256$, and the grid radii were 300, 30 and 1 kpc. Effective
gravity softening $\epsilon$ in particle--mesh codes is equal to
half of the grid cell \citep{J11}, so in the centre within 1 kpc
$\epsilon \approx 4$\,pc, while outside the centre in the disc
$\epsilon \approx 117$\,pc.

For the largest simulations in our set of runs we use the modification
of the recently developed and very popular $N$-body Tree-GPU code implementation
{\tt bonsai2} \citep{BGPZ2012a, BGPZ2012b}, which includes expansion for force computation up to quadrupole order. The code was used with  very
conservative opening angle ($\theta$ = 0.5) and with the individual
gravitational softening option (the softening was set to 10 pc). In our
runs we use the simple leap-frog integration scheme with a fixed time-step
$\Delta$t = 0.2~Myr to advance the particle positions and velocities
during the calculation.

For the simulations of the rigid halo/bulge models and to analyse the snapshot data of all simulations we also use the self coded Tree-GPU based
gravity calculation routine
{\tt ber-gal0}\footnote{\tt ftp://ftp.mao.kiev.ua/pub/users/berczik/ber-gal0/}
\citep{ZBGPJ2015}, which includes expansion for force computation up to monopole order, with the same ($\theta$ = 0.5) opening angle.

In the last years we have already used and extensively tested our hardware
accelerator based on the tree gravity calculation routine in a few
of our galactic dynamics projects and get quite accurate results with
a good performance \citep{SB2005, BJBB2008, PGBSD2010, PGBCS2011}.

The current set of simulations was carried out with the GPU version of
the code using local GPU clusters available at the authors' institutions
(ARI: {\tt kepler}, MAO: {\tt golowood}, NAOC: {\tt laohu}) and also the
specially dedicated for SFB 881 (``The Milky Way System'') GPU cluster
{\tt MilkyWay}, located in the J\"ulich Supercomputing Centre in Germany.

\subsection{Live disc/bulge/halo runs}

Tab.\,\ref{tab_runs_live} contains parameters of live runs ,
where all components, disc, bulge and halo, are live, for the basic
model. Capital letters denote the numerical code: `S' for {\tt
Superbox} and `B' for {\tt bonsai2}. Most of the runs have equal
mass of the particles within a component, but two runs denoted by
`m' have multimass halo particles in order to achieve a better
resolution in the bar region. For this, we modified the GalactICS
code utilising the same strategy as described in \citet{DBS09}. In
the region between 0.1 and 1 kpc, the number density ratio of our
multimass and single mass runs varies from 10 to 100, thus the
effective numerical resolution there is enhanced by this factor.

\begin{table*}
\begin{center}
\begin{tabular}{l l l l l l l l l}
\hline
 Run & $N_\textrm{tot}$ & $N_\textrm{d} | N_\textrm{b} | N_\textrm{h}$  & $\epsilon$ & e-period &  $\Omega_\textrm{p}$ & $\omega_\textrm{I} [B]$ & $\omega_\textrm{I}$ [E]\\
\hline
S1  & 5.6   & $ 1.1 | 0.5 | 4.0$   & 4--117 & $0.9<t<1.5$ & 50.6 & 3.83 & $ 3.66 \pm 0.23 $ \\ 
S3m & 16.75 & $ 6.0 | 1.5 | 9.25$  & 4--117 & $0.9<t<1.5$ & 52.2 & 3.59 & $ 3.65 \pm 0.25 $ \\[1mm]  
B1  & 5.6   & $ 1.1 | 0.5 | 4.0$   & 10 & $0.1<t<1.0$ & 54.6 & 4.18 &  $ 4.07 \pm 0.29 $ \\ 
B2m & 16.75 & $ 6.0 | 1.5 | 9.25$  & 10 & $0.8<t<1.3$ & 54.4 & 4.33 &  $ 4.46 \pm 0.33 $ \\ 
B3  & 104.5 & $ 6.0 | 1.5 | 97  $  & 10 & $0.9<t<1.5$ & 54.9 & 4.36 &  $ 4.58 \pm 0.19 $ \\ 
\hline
\end{tabular}
\end{center}
\vspace{-2mm}
\caption{Parameters of live runs of the basic model. Total number of particles $N_\textrm{tot}$ and number of particles in components are given in millions (M); gravity softening parameter $\epsilon$ is in pc; e-period -- in Gyr; pattern speeds $\Omega_\textrm{p} = \omega_\textrm{R}/2$ and growth rates $\omega_\textrm{I}$ are given in km/s/kpc $\approx$ Gyr$^{-1}$.  `m' denotes multi mass setup for halo stars.  }
\label{tab_runs_live}
\end{table*}

The total number of particles varies from 5.6M to 104.5M. Our
default runs have 16.75M, with 6M particles in the disc, 1.5M in the
bulge, and 9.25M in the halo. Runs with smaller and higher number of
particles are used to show the effect of an $N$-variation. Mass of
halo particles in the B3 run is only twice heavier than the
disc and bulge particle mass, so this run is used to show the
absence of disc heating from heavier halo particles due to shot
noise.

All runs initially show a minor non-equilibrium resulting in a rapid
change of the disc thickness and radial redistribution of the
central part. So, an average disc height $(\langle z^2
\rangle)^{1/2}$ within $R_\textrm{d}$ is equal to 274 pc at
$t=0$, but it is already 290 pc at $t=10$ Myr. Both radial and
vertical oscillations disappear during one typical Jeans time $\sim$
100 Myr. This is a new equilibrium state.

\begin{figure}
   \centerline{\includegraphics [width = 85mm]{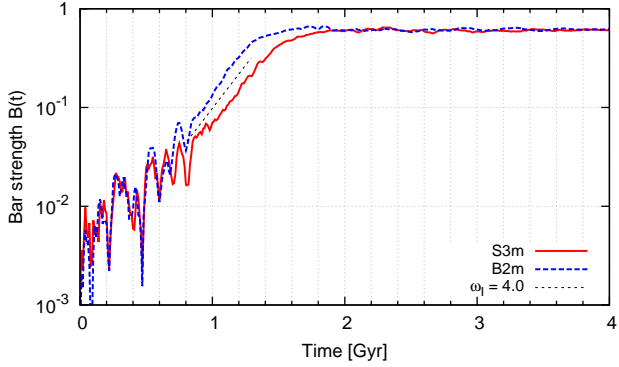}}
   \caption{Bar strength in the default live runs S3m and B2m. The dotted line shows growth of the bar strength with rate $4$ Gyr$^{-1}$.}
   \label{fig:bs_live}
\end{figure}

The bar triaxiality parameters of $I_{yy}/I_{xx}$ and
$I_{zz}/I_{xx}$, where $I$ are components of the inertia ellipsoid,
\begin{equation}
I_{xx} = \sum\limits_{j} m_j x^2_j\ ,\quad I_{yy} = \sum\limits_{j} m_j y^2_j\ ,\quad I_{zz} = \sum\limits_{j} m_j z^2_j\ ,
\label{eq:tri}
\end{equation}
(with the long axis of the bar oriented along the x-axis)
provide the simplest tool to analyse the disc evolution and
stability. Here $m_j$ denotes the mass of disc particles, and we
assume that $j$ spans particles within the characteristic scale
length $R_d$. In Fig.\,\ref{fig:bs_live} we show the bar strength
defined as
\begin{equation}
B(t) = 1 - I_{yy}/I_{xx}
\label{eq:barstr}
\end{equation}
in linear-log axes for S3m and B2m runs. As a rule, such curves consist of three parts: low amplitude lag, the linear regime growth phases, and plateaus. The latter two are clearly interpreted as exponential growth (its period is indicated as `e-period' in Tab.\,\ref{tab_runs_live}), and saturation of the instability. Note that blue and red curves follow each other even in deep oscillations until $\approx 0.8$ Gyr.

The lag part can be short, as in the case of run B1, or last about 1
Gyr (this is the case for the others). During this time, the bar
strength increases and decreases, perhaps showing a tendency to
increase, but not exponentially. The lag for exponentially fast bar
formation can be explained by (i) superposition of modes with
different frequencies and azimuthal numbers, or (ii) Poisson noise
that operates in stellar systems and affects bar formation. The same
lag can be found in \citet{DBS09}, Fig. 15, where one can observe a
difference between multi-mass 100M run and single mass 100M run (red
and black curves respectively). Note that the number of disc
particles and initial noise level in these two runs was the same.

\begin{figure*}
\centerline{
\includegraphics [width = 200mm]{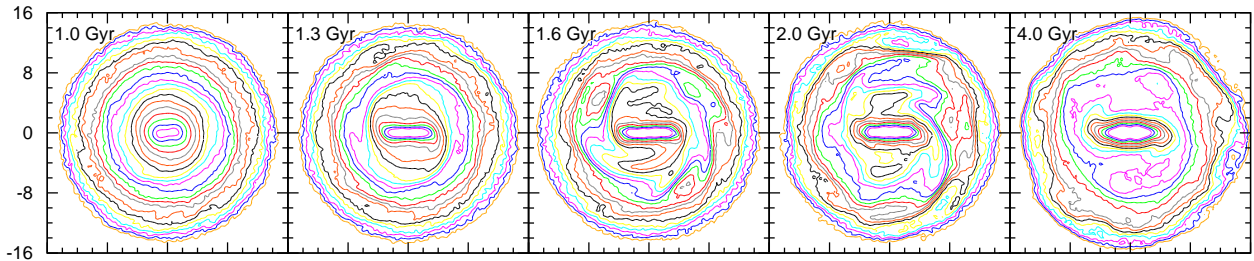}
} \caption{B2m bar patterns oriented along the x-axis at
different stages of bar evolution. The curves are isolines of the
density evenly spaced in log scale (10 levels for every factor of
10). Each frame size is 32x32 kpc. } \label{fig:gS1}
\end{figure*}

After the lag, an instability begins to operate and forms a bar.
Fig.\,\ref{fig:gS1} shows snapshots at times 1.0, 1.3, 1.6, 2.0, and
4 Gyr of the bar oriented along the $x$-axis.

Pattern speed $\Omega_\textrm{p}$ can be calculated from the change
of azimuthal phase of the bar. When the bar amplitude is low, the
bar phase is uncertain, so the pattern speed is determined with
large errors. During the linear growth, the pattern speed is
constant (these values are given in  Tab.\,\ref{tab_runs_live}).
After the instability saturates, we observe a gradual decrease of
$\Omega_\textrm{p}$. This effect is known as {\it bar braking}.

To quantify the bar shape we fit ellipses to the isophotes and associate a bar radius, $R_\textrm{b}$ with the radius at which ellipticity $\varepsilon (r)\equiv 1-b_e/a_e$ ($a_e$ and $b_e$ are a major and minor semiaxes of ellipses) declines $\sim$ 15\,\% from its maximal value \citep{Inma}. For $t=1$ Gyr, the bar radius is about 1.9 kpc, and $\varepsilon \simeq 0.6$. At the end of the linear growth, $t\approx 1.3$, the bar radius is 4 kpc, with ellipticity  0.72. Some other values are given in Tab.\,\ref{tab_rb_live}.

\begin{table}
\begin{center}
\begin{tabular}{l l l l l l l l}
\hline
Time [Gyr] & 1.0  & 1.3 & 1.6 & 2.0 & 2.5 & 3.0 & 4.0 \\
\hline
$R_\textrm{b} $ [kpc]  & 1.9 &4.0 &4.9 &3.9 &2.6 &1.8 &1.6  \\
$\varepsilon $  & 0.59 &0.72 &0.74 &0.74 &0.76 &0.77 &0.76  \\
\hline
\end{tabular}
\end{center}
\vspace{-2mm} \caption{Parameters of the bar (radius, ellipticity)
in B2m run at different moments in time.} \label{tab_rb_live}
\end{table}

\begin{figure}
\centerline{\includegraphics [width = 85mm]{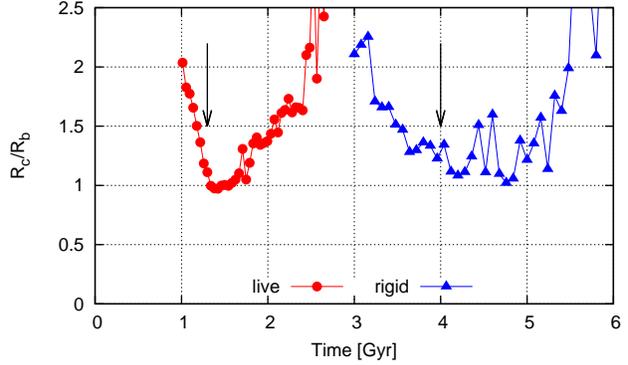}}
\caption{Ratios of the corotation radius $R_\textrm{c}$ to the bar radius $R_\textrm{b}$ for the default live B2m and default rigid T1r runs. Arrows show the adopted time of saturation of bar instability.}
\label{fig:rb}
\end{figure}

A ratio ${\cal R}$ of the corotation radius $R_\textrm{c}$ to the
bar radius $R_\textrm{b}$ is used to distinguish between fast and
slow bars. For usual fast bars, corotation is not far beyond the
bar's end, $1 \leq {\cal R} \leq 1.4$ \citep{DS00}. According to
Fig.\,\ref{fig:rb} which shows this ratio as a function of time,
${\cal R}$ reaches one at $t\approx 1.3$ Gyr, and stays there up to
$t\approx 1.6$ Gyr. Since the bar slows down after $t\approx 1.3$
Gyr, and thus the corotation radius increases, the bar increases in
length during this period and reaches 5 kpc. Afterwards, it begins
to shorten.

The growth rate can be determined from the slope of the bar strength, $B(t)$, in the linear growth phase. In Tab.\,\ref{tab_runs_live}, this quantity is denoted as $\omega_\textrm{I} [B]$. This is close to, but not coincident with the growth rate of the $m=2$ wave amplitudes, which can be measured using a surface density wave amplitude,
\begin{equation}
    \tilde \Sigma (R) = \frac 1{\delta M_\textrm{d}(R)} \sum_{j} m_j \textrm{e}^{-im\theta_j}\ ,
\label{eq:sd1}
\end{equation}
and radial and azimuthal velocity wave amplitudes:
\begin{align}
    \tilde V_R (R) & = \frac 1{\delta M_\textrm{d}(R)} \sum_{j} m_j v_{R,j} \textrm{e}^{-im\theta_j}\ , \label{eq:vr1}\\
    \tilde V_\theta & = V_\theta - V_c \tilde \Sigma (R)\ , \label{eq:vp1} \\
    V_\theta (R) & = \frac 1{\delta M_\textrm{d}(R)} \sum_{j} m_j v_{\theta,j} \textrm{e}^{-im\theta_j}\ , \label{eq:vp2}
\end{align}
where $V_c=R\Omega(R)$ is the circular velocity, $\delta
M_\textrm{d}(R) = \sum_{j} m_j$ is a mass of particles in a ring of
width $\delta R$ near radius $R$. Calculation of growth rates from
$N$-body simulations shows $R$-dependence, but it is weak for the
particles within the central part of the disc where one global
mode dominates. For such cases, we use a mass weighted radial
average of these amplitudes in the rings:
\begin{align}
    E(\cdot)  & = \frac1{M_\textrm{d}(R_\textrm{d})} \sum \delta M_\textrm{d}(R) |\cdot| \ ,
\label{eq:ave}
\end{align}
where $M_\textrm{d}(R_\textrm{d}) = \sum \delta M_\textrm{d}(R)$ is
a total mass of disc particles within $R_\textrm{d}$. The mass
weighted radial average for $\tilde \Sigma (R)$ gives the usual
$A_2/A_0$ ratio frequently used to characterise the bar mode
amplitude \citep[see, e.g.,][]{S16}.

\begin{figure}
\centerline{\includegraphics [width = 85mm]{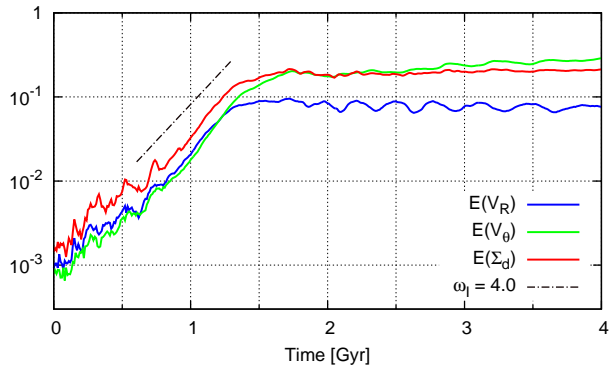}}
\caption{Growth of the wave amplitudes in B2m run. Dash-dotted line shows growth of the bar strength with rate $4$ Gyr$^{-1}$. Wave amplitudes are given in N-body units.}
\label{fig:bp07}
\end{figure}

Wave amplitudes $E(\tilde \Sigma_\textrm{d})$, $E(\tilde V_R)$, and
$E(\tilde V_\theta)$ are shown in Fig.\,\ref{fig:bp07}. Comparison
with Fig.\,\ref{fig:bs_live} shows almost the same linear growth
from 0.8 to 1.3 Gyr, but smoother behaviour of the wave amplitudes
at $t<0.8$ Gyr. The growth rates $\omega_\textrm{I}$ inferred from
each of the components slightly differ. E.g., for the surface
density we obtained 4.33, while for the radial and azimuthal
velocity amplitudes they are 4.22 and 4.84 Gyr$^{-1}$, respectively.
In all runs, $\omega_\textrm{I}$ determined from $E(\tilde
\Sigma_\textrm{d})$ and $E(\tilde V_R)$ are close (deviation is 2
per cent), while $\omega_\textrm{I}$ inferred from  $E(\tilde
V_\theta)$ is from 7 to 15 per cent higher. An average value of
these three values, along with the corresponding standard deviation,
is given as $\omega_\textrm{I} [E]$ in Tab.\,\ref{tab_runs_live}.

\begin{figure}
\centerline{\includegraphics [width = 85mm]{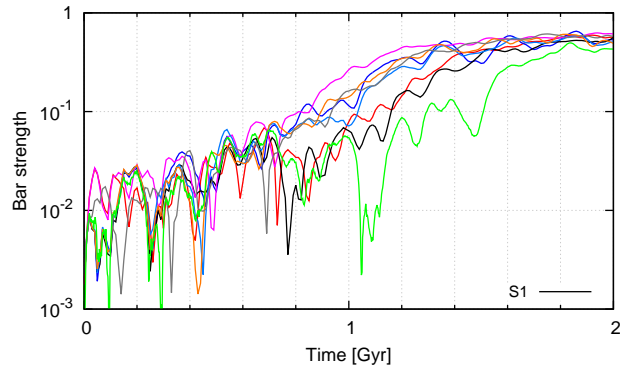}}
\caption{Bar strength curves for S1 model (8 runs which differ insignificantly). }
\label{fig:bs_s1}
\end{figure}

The superbox frequencies ($\Omega_\textrm{p}$ and
$\omega_\textrm{I}$) are smaller on average compared to the tree
code runs. This can be due to somewhat larger gravity softening of
superbox models, since softening makes gravity interaction weaker,
and thus one would expect lower frequencies for eigen-oscillations
\citep{EP13}. Another explanation can be connected with
stochasticity inherent to N-body simulations. \citet{SD09} note that
small changes of insignificant parameters can lead to significant
changes in galaxy evolution after bar formation. For the S1 run we
varied disc inclination relative to the grid, grid flattening, and
also made minor changes in distributing particles over the phase
space (8 runs including S1 itself). As is seen in
Fig.\,\ref{fig:bs_s1}, the bar strength curves begin to rise at
different times with mean lag $\sim$ 0.6 Gyr.

Note that all curves have a low amplitude lag in the beginning, and
some even lack the definite linear growth phase. Special methods
such as quiet start \citep{SA86} are used to decrease initial
nonaxisymmetric perturbations in order to obtain clear linear
growth. We presume that shot noise at some level is an essential
ingredient of galaxy dynamics, and thus here we deliberately avoid
such procedures as artificial.

The given above analysis and results of others
\citep[e.g.,][]{WPD08} leave no doubt that bar instability takes
place in numerical bars in the presence of ILR (for $\Omega_\textrm{p}=55$ km/s/kpc, ILR radius is $\sim 0.4$ kpc), and so discrepancy
with the theory needs to be explained. 

\begin{figure}
\centerline{\includegraphics [width = 85mm]{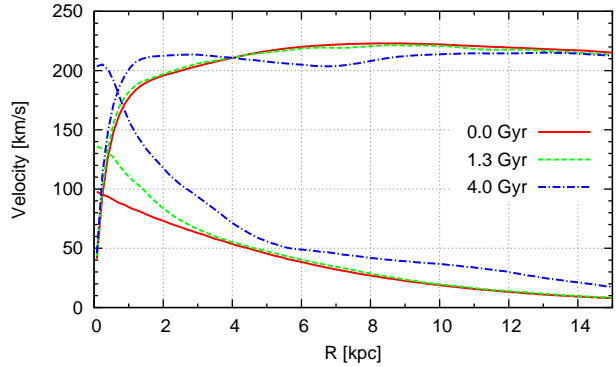}}
\caption{Circular velocity and radial velocity dispersion profiles of the B2m run at 0, 1.3, 4 Gyr. }
\label{fig:vcsr}
\end{figure}

Axisymmetric properties of the disc do not change until the end of
the growth phase, as can be seen in Fig.\,\ref{fig:vcsr}. In the end
(1.3 Gyr), the velocity dispersion is affected only within a 2 kpc
radius, while the circular velocity is almost preserved, as well as
the axisymmetric surface density background. Thus during the growth
phase, the stability parameter $Q$ remains
unchanged, and one can expect that the main characteristics of the
instability (pattern speed, growth rate, shape of the patterns) can
be reproduced from linear perturbative analysis. For the moment,
however, we are able to calculate linear global modes only for
razor-thin discs embedded into a rigid bulge and halo. So, galaxy
models with rigid spheroidal components have to be considered.

\subsection{Rigid halo/bulge runs}

A series of runs with rigid spheroidal components denoted by `r' is
presented in Tab.\,\ref{tab_runs_rigid}. Here `S' again denotes
Superbox runs and `T' denotes {\tt ber-gal0} Tree code runs.
The rigid components were modelled in three ways.

\begin{figure*}
\centerline{
\includegraphics [width = 200mm]{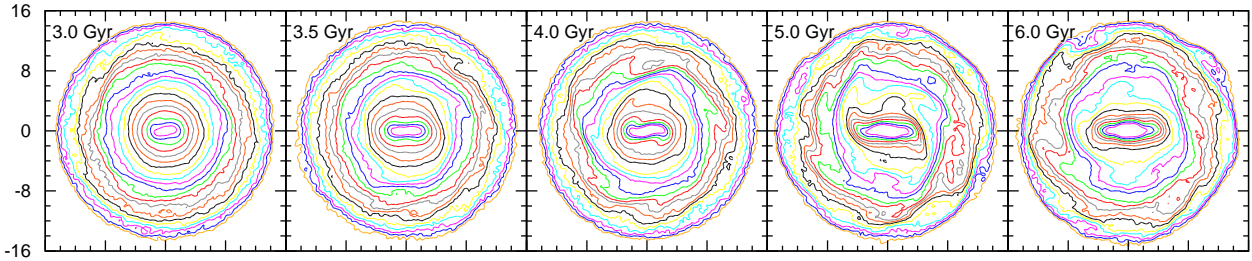}
}
\caption{Same as in Fig.\,\ref{fig:gS1} for the default rigid run T1r.}
\label{fig:gS2}
\end{figure*}

\begin{table*}
\begin{center}
\begin{tabular}{l l l l l l l l l}
\hline
 Run & $N_\textrm{tot}$ & $N_\textrm{d} | N_\textrm{b} | N_\textrm{h}$  & $\epsilon$ & e-period &  $\Omega_\textrm{p}$ & $\omega_\textrm{I} [B]$ & $\omega_\textrm{I}$ [E]\\
\hline
S2r (FP) & 5.6   & $ 1.1 | 0.5 | 4.0$   & 4--117 & $2.1<t<3.5$ & 47.8 & 1.06 & $ 1.09 \pm 0.10 $   \\ 
S4r (AX) & 1.1   & $ 1.1 | - | - $      & 12     & $1.0<t<2.5$ & 51.1 & 1.21 & $ 1.12 \pm 0.04 $   \\ 
S5r (SP) & 1.1   & $ 1.1 | - | - $      & 12     & $1.5<t<3.0$ & 50.2 & 1.84 & $ 1.51 \pm 0.13 $   \\[1mm] 
T1r (FP) & 16.75 & $ 6.0 | 1.5 | 9.25$  & 10     & $2.3<t<3.6$ & 51.5 & 1.77 & $ 1.81 \pm 0.06 $   \\ 
T2r (SP) & 6.0   & $ 6.0 | -|-$         & 10     & $2.4<t<3.8$ & 54.3 & 1.2  & $ 1.10 \pm 0.03 $   \\ 
T3r (AX) & 6.0   & $ 6.0 | -|-$         & 10     & $2.2<t<3.5$ & 52.0 & 1.86 & $ 1.90 \pm 0.07 $   \\ 
\hline
\end{tabular}
\end{center}
\vspace{-2mm}
\caption{Same as in Tab.\,\ref{tab_runs_live}, for the rigid runs of the basic model. }
\label{tab_runs_rigid}
\end{table*}

The simplest but expensive way to simulate galaxies with rigid components is to fix coordinates of bulge and halo particles in space, and use them in calculation of the potential. These runs are S2r and T1r, marked by `(FP)'.

In runs marked by `(SP)' (S5r and T2r), halo and bulge potentials
are approximated by spherically symmetric functions. The forces from
the halo in this case can be calculated analytically, given the NFW
potential. The forces from the bulge can be obtained through
interpolation from the forces tabulated by integrating density that
is given analytically. This is the cheapest but inaccurate way,
because in the presence of the disc, distributions of particles in
the bulge and halo become axisymmetric rather than spherically
symmetric. This manifest itself in larger radial and vertical
initial oscillations (e.g., initial disc height jumps from 274 to
305 pc), which are damped after the first $\sim$ 150 Myr.

Finally, in runs marked by `(AX)' (S4r and T3r), halo and bulge
potentials are axisymmetric. The potentials and forces from halo and
bulge have been tabulated from the actual distribution in the
meridian $(R,z)$ plane, and interpolated linearly to positions of
disc particles. The initial height jump in this case is the same as
in live and (FP) rigid runs, i.e. from 274 to 290 pc.

The superbox runs were made with $N_\textrm{d}=1.1\cdot 10^6$. The
grid sizes for the S2r were 1, 30, 300 kpc, but in S4r and S5r we
used another set of grids, 3, 9, 27 kpc, since now there is no need
for grids with halo size. Therefore, the effective softening for
these runs is only 12 pc. For tree code runs, we used
$N_\textrm{d}=6\cdot 10^6$ particles, with the standard softening
$\epsilon=10$ pc.

Fig.\,\ref{fig:gS2} shows snapshots of the bar in T1r run (used as a
default) oriented along the $x$-axis at times 3, 3.5, 4, 5, and 6
Gyr.

\begin{figure}
\centerline{\includegraphics [width = 85mm]{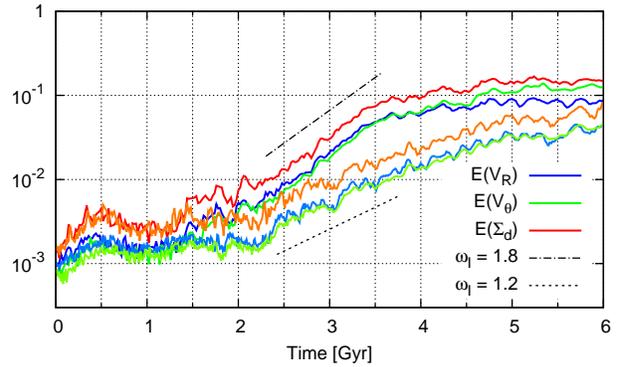}}
\caption{Same as in Fig.\ref{fig:bp07} for the rigid runs T1r (upper group of lines -- blue, green, and red) and T2r (lower group of lines -- orange, light green, light blue). Dash-dotted and dotted lines show growth with rates 1.8 and 1.2 Gyr$^{-1}$.}
\label{fig:bp12}
\end{figure}

Rigid runs show more extensive lag before the exponential bar growth
($\approx$ 1.9 Gyr on average) than the live runs ($\approx$ 
0.6 Gyr). During the lag, the bar strength (Eq. \ref{eq:barstr})
and wave amplitudes (Eqs. \ref{eq:sd1}--\ref{eq:vp2}) are not
necessarily constant, but rather rise and fall, as shown in
Fig.\,\ref{fig:bp12}. Note also that the end of bar formation is
more uncertain in the rigid case. For T1r it can be associated with
3.5 or 4 or even with 5 Gyr, although the linear growth ends at 3.5
Gyr.

The pattern speeds $\Omega_\textrm{p}$ determined during the
e-period (given as `e-period' in the Tab.\,\ref{tab_runs_rigid})
scatter from 47.8 to 54.3 km/s/kpc. Again, $\Omega_\textrm{p}$ is
lower for the Superbox code. After the linear growth, the bar slows
down in the same manner as in the live runs.

The ratio ${\cal R} = R_\textrm{c}/R_\textrm{b}$ for the T1r run is
shown at Fig.\,\ref{fig:rb} with blue triangles. Compared to the B2m
live run, it is more noisy on minimal values, and the minimum is
higher, $\sim$ 1.2. The latter is both because the bar is shorter,
and the pattern speed is smaller, resulting in larger corotation
radius. At the end of bar formation, assumed at $t=4$ Gyr, the bar
radius is 3.8 kpc and ellipticity is $\varepsilon = 0.65$. The
maximum radius of 4.3 kpc is achieved at 4.8 Gyr (ellipticity 0.68).
Some other values are given in Tab.\,\ref{tab_rb_rigid}.

\begin{table}
\begin{center}
\begin{tabular}{l l l l l l l l}
\hline
Time [Gyr] & 3.0  & 3.5 & 4.0 & 4.5 & 5.0 & 5.5 & 6.0 \\
\hline
$R_\textrm{b} $ [kpc]  & 1.8 &3.0 &3.8 &4.1 &4.1 &2.8 &1.7  \\
$\varepsilon $  & 0.55 &0.63 &0.65 &0.67 &0.70 &0.72 &0.73  \\
\hline
\end{tabular}
\end{center}
\vspace{-2mm}
\caption{Parameters of the bar (radius, ellipticity) in T1r run at different moments of time (spline smoothed values).}
\label{tab_rb_rigid}
\end{table}

The growth rates $\omega_\textrm{I}$ determined during the e-period
spread in the interval from 1.06 to 1.9 Gyr$^{-1}$. For each run,
the values inferred from the bar strength (Eq. \ref{eq:barstr}) and
the wave amplitudes (Eqs. \ref{eq:sd1}--\ref{eq:vp2}) are generally
in agreement. T1r run, in which rigid components are modelled by
fixed particles, gives almost the same frequencies as the T3r run,
where halo and bulge are represented by axisymmetric potentials, but
different from T2r. Superbox runs are similar: fixed particle
(FP) and axisymmetric (AX) runs are close, but the spherically
symmetric one is different. One possible reason is that the
spherically symmetric runs adjust to a new equilibrium that is
different from one in the FP and AX runs. Another, more likely
reason is that all runs are subject by stochasticity, which affects
not only the lag, but the frequencies as well. Comparison with live
runs shows that the stochasticity effect is stronger for less
unstable models.

Lower growth rates in the rigid runs are not expected intuitively.
One would expect a priori that spherical isotropic (i.e. stable)
halos cannot affect disc stability, since the ratio $\rho/\sigma^2$
for the disc component is much larger than the same ratio for the
halo \citep{MS74}. However, the growth rates in runs with live and
rigid halos differ by a factor of 2, or even larger.  Indeed, while
the volume density of halo particles is low in the disc, and their
dispersion is high, resonance halo particles are not confined to the
disc. Instead, they occupy large resonance regions above and below
the disc, and their number and influence turn out to be significant.
Considerably weaker bars in the rigid halos were also obtained by
\citet{A02}, and explained by the additional interaction of the bar
with halo particles, mainly on the corotation resonance.

\section{Global mode analysis}

The confrontation of $N$-body simulations and linear perturbation
global mode analysis is a very interesting but challenging problem.
As we already mentioned above, direct application of the matrix
methods fails in cuspy models because of the presence of an ILR and
inability to take into account live halos. Thence, one can try to
reproduce modes in $N$-body models with rigid halos, particularly
since the pattern speeds are not affected by replacing the live halo
with the rigid one.

For bar formation in razor thin discs, it is crucial to have the
precession curve $\Omega_\textrm{pr}(R) \equiv
\Omega(R)-\kappa(R)/2$ with a maximum. If the maximum is not too
high, self gravity of the disc is able to support bar-like
perturbations with pattern speeds above the maximum \citep{EP04}.
This is not the case in cuspy models, where the profile grows
infinitely as $r^{-\alpha/2}$ in the centre. One should note
however, that $N$-body discs are different from discs considered in
linear perturbation analysis, first of all because  randomly
distributed particles are used to describe the stellar components
leading to different stochastic effects. Besides, the potential
issues are numerical and depend on equilibrium accuracy, cusp
resolution, gravity softening, disc thickness, and the form of the
distribution function (DF). All these may distort the precession
curve and affect the location of the ILR, important to the appearance of
global modes.

\subsection{Numerical accuracy}

\begin{figure}
\centerline{\includegraphics [width = 85mm]{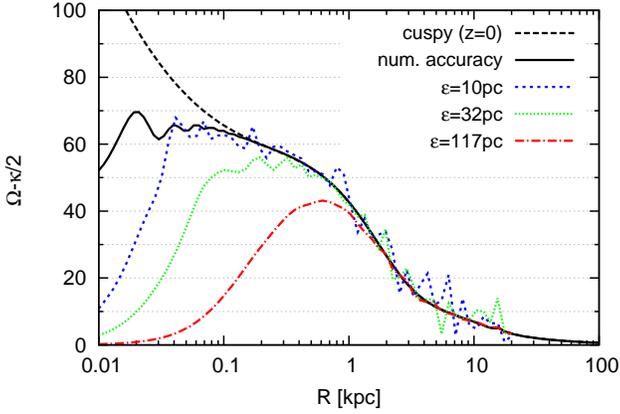}}
\caption{The $\Omega_\textrm{pr}$ curves affected by numerical
accuracy and softening. The black dashed line shows the true cuspy
curve for particles in the equatorial plane, $z=0$; the black solid
line shows the curve affected by insufficient numerical accuracy;
colour lines show the distortion of the black solid line due to gravity
softening. } \label{fig_soft}
\end{figure}

The evaluation of the epicyclic frequency $\kappa$ requires the
calculation of a second derivative of the potential:
\begin{equation}
\kappa^2 = 3\Omega^2 + \Phi''\ .
\label{eq:epf}
\end{equation}
If the potential in a particle generator is obtained from some
iteration procedure, then $\kappa(R)$ can differ significantly from
the cuspy law $\propto r^{-\alpha/2}$. In fact, setting a grid
parameter in GalactICS to ordinary $\Delta R = 10^{-2}$ one can
resolve $\Omega_\textrm{pr}(R)$ correctly only until a radius of 0.2
kpc. On smaller radii an artificial maximum of 70 km/s/kpc appears
(black solid line in Fig.\,\ref{fig_soft}). To get rid of the
maximum at least in the range $\Omega_\textrm{pr} < 100$ km/s/kpc,
it is necessary to set this parameter sufficiently small, e.g.
$\Delta R = 10^{-3}$ (black dashed curve).

\subsection{Cusp resolution}

The number of bulge particles $N_\textrm{b}$ can be insufficient to
reproduce a cuspy $\Omega_\textrm{pr}$ profile. To derive an
estimate for a sufficient number of particles, we note that the main
uncertainty comes from the calculation of $\kappa$. For spherically
symmetric systems, $\kappa^2 = 4\pi G \rho (r) + \Omega^2$, thus
$$
\frac{\delta \Omega_\textrm{pr}}{\Omega_\textrm{pr}} \sim
 \frac{8}\alpha  \frac{\delta \kappa} {\kappa} \sim \frac{4}\alpha \frac{\delta \rho_\textrm{b}} {\rho_\textrm{b}} \ .
$$
The uncertainty of the bulge density $\delta \rho_\textrm{b}/{\rho_\textrm{b}}$ can be estimated as $(\delta N)^{-1/2}$, where $\delta N$ is the number of particles in a shell of radius $r$ and width $\delta r$, for small radii  $\delta N \sim 4\pi N_\textrm{b} (r^2/R^3_e) \delta r $. If we require the epicyclic frequency to be accurate within a small value $\varepsilon'$, the number of particles in the shell should be
\begin{equation}
N_\textrm{b} \left(\frac{r}{R_e}\right)^{3} \frac {\delta r}{r} \gtrsim (\varepsilon'\alpha )^{-2}\ .
\label{eq:nbe}
\end{equation}
For example, for $\varepsilon' = 0.1$ accuracy, to resolve
$\Omega_\textrm{pr}(R)$ in our model up to $0.1$ kpc, it is
sufficient to have 0.2M particles. However, to resolve a radius of
0.05 kpc one needs at least 1.5M particles.

\subsection{Gravity softening}

A change of the gravity interaction law from the Newton law $r^{-1}$
to a Plummer law $(r^2 + \epsilon^2)^{-1/2}$ in Tree codes or the
usage of meshes in the particle--mesh code leads to a smearing of
the cusp in the density distribution, and consequently to a
flattening of the precession profile, i.e. its shape becomes typical
for galaxies with cores. These galaxies have finite density and thus
obey the solid body rotation law, $\Omega \approx \textrm{const}$,
and vanishing $\Omega_\textrm{pr}$ in the centre.
Fig.\,\ref{fig_soft} shows three examples of $\Omega_\textrm{pr}$
calculated for the initial particle distribution in the B2m run,
which are affected by 10, 32 and 117 pc softening. For $\epsilon =
117$ pc, the obtained profile is 10 per cent lower than the cuspy
one already at 1 kpc. For $\epsilon = 32$ pc, the
$\Omega_\textrm{pr}$ falls off the correct profile at $R\approx 0.5$
kpc, and peaks at 0.2 ... 0.3 kpc. The value of $\epsilon = 10$ pc
adopted in our Tree code simulations alters the profile given by the
black solid curve only within $R<0.03$ kpc.

In our calculations we use $N_\textrm{d}=1.1$M and 6M particles to
represent the disc component, and $N_\textrm{b}=0.5$M and 1.5M
particles to represent the bulge component. A natural condition for
the softening is that softening volume should contain several
particles to properly suppress both potential fluctuations and
2-body scattering at short range. For disc particles $a^3
N_\textrm{d} \sim (z_\textrm{d} R^2_\textrm{d})$, which gives for a
mean distance between particles $a \sim  (z_\textrm{d}
R^2_\textrm{d} /N_\textrm{d} )^{1/3} = 8 ... 14$ pc. An estimate for
the bulge gives $a \sim (R^3_e /N_\textrm{b} )^{1/3} = 6 ... 11$ pc.
Thus, our choice of $\epsilon = 10$ pc is sufficiently large for the
default runs, but may be a source of some relaxation in models with
a smaller number of particles.

Large values of the gravity softening parameter $\epsilon$ should
also affect the equilibrium, since GalactICS lacks this parameter.
We expect, however, that it happens when  $\epsilon$ becomes a
noticeable fraction of the characteristic scales, which are of the
order of 1 kpc.

\subsection{Thickness}

In contrast to razor thin discs used in linear perturbation
analysis, real discs and their $N$-body counterparts are
three-dimensional. While motion in the plane is always integrable,
in the sense that the Hamilton--Jacobi equation separates leading to
periodic radial and azimuthal motion, this is not generally the case
in 3D motion. It is likely that the orbits of many disc
particles remain regular, i.e. they respect three integrals of
motion and have three fundamental frequencies. However, direct
integration of orbits shows that for particles with $R \lesssim z
\sim z_d$, the radial frequency is no longer fundamental, but rather a
combination of frequencies. So, the radial frequency $\Omega_{R}$, as
well as the usual ILR, is not defined at $R \lesssim z_d$. Certainly,
for larger radii, $R \gg z_d$, the motion is nearly flat, and usual
frequencies can be introduced. For example, for nearly circular
orbits in the $z=0$ plane the vertical motion can be separated so that
particles oscillate in a `vertical' potential $\Phi(R,z) -
\Phi(R,0)$. However, in the case $R \lesssim z_d$ the ILR should be
substituted by a resonance with a different combination of
frequencies, and interaction with waves on this resonance should be
reconsidered.

\begin{figure}
\centerline{\includegraphics [width = 85mm]{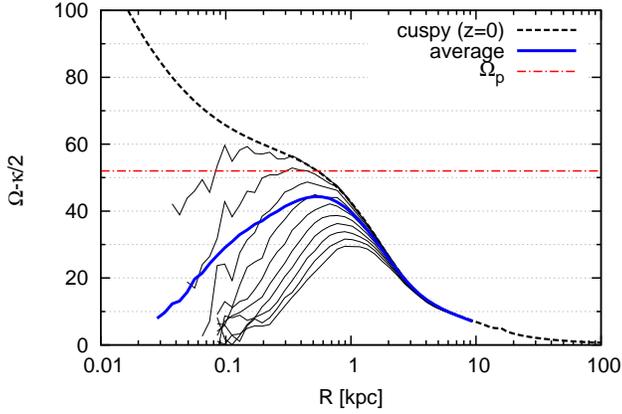}}
\caption{The precession profiles of the B2m initial distribution
calculated for different elevations of particles above the plane
(each thin solid curve is calculated for particles in the layer
$\Delta z = 50$ pc width, starting from $z=0$), and average
$\overline \Omega_\textrm{pr}$ (thick solid curve). The dashed line
shows the cuspy profile for particles in the the equatorial plane.
The dash-dotted line shows the bar pattern speed obtained in rigid
halo models.} \label{fig_thick}
\end{figure}

To proceed further, we consider a simple model that allows replacing the complex 3D orbital motion with planar motion. Only a small fraction of particles that travel close to the galactic symmetry plane $z=0$ feels the cuspy form of the $\Omega_\textrm{pr}$ profile. Angular velocity $\Omega$ and epicyclic frequency $\kappa$ out of that plane do not have a cuspy singularity $r^{-\alpha/2}$, since
\begin{equation}
    \Omega^2(R,z) = \frac{1}{R}\frac{\p \Phi(R,z)}{\p R}\ ,
\end{equation}
and $\Phi(R,z)$ is smooth at $R = 0$ if $z\ne 0$. Thus nearly all
particles move as if there is a cored rather than cuspy
$\Omega_\textrm{pr}$ profile. Fig.\,\ref{fig_thick} shows precession
curves calculated for different height $z$ of the initial
distribution of the B2m run (each thin line represents
$\Omega_\textrm{pr}(R)$ calculated for particles in a flat layer
with a thickness of 50 pc), and the averaged profile
\begin{equation}
    \overline \Omega_\textrm{pr} = \overline \Omega -\overline \kappa/2\ ,
    \label{eq:ave_ompr}
\end{equation}
where overlines denote square averages, i.e. $(\overline \Omega)^2 =
\langle \Omega^2 \rangle_z$ and $(\overline \kappa)^2 = \langle
\kappa^2 \rangle_z \equiv 4\langle \Omega^2 \rangle_z + d \langle \Omega^2 \rangle_z /dr $. The maximum of the averaged curve turns out to
be 43.7 km/s/kpc at 0.52 kpc, which is well below the pattern speed
obtained in our simulations with rigid halos. The effective
precession curve roughly coincides with the curve calculated for
particles that are $z\sim 150 ... 200$ pc out of the plane.

\subsection{The Distribution function}

The disc phase space distributions used in GalactICS with the form
$f_0(R,z, v_R, v_\theta, v_z)$ cannot be reproduced by DFs with the
form $F_0(E,L_z)$ or $F_0(J_r, L_z)$ used in matrix methods, since
they do not account for the vertical motion. Even if one neglects
the vertical motion and considers the Schwarzschild DF
\begin{equation}
    f_0(R, v_R, v_\theta) = \left.\frac{\Omega}{\pi\kappa} \frac{\Sigma_\textrm{d}}{\tilde\sigma^2_R}\right|_{R} \exp \left( -\frac{v^2_R + \gamma^2 \tilde v^2_\theta}{2\tilde\sigma_R^2(R_c)}\right)\ ,
    \label{eq:dfv}
\end{equation}
(here $\tilde v_\theta \equiv v_\theta - R\Omega(R)$, $\gamma \equiv
2\Omega/\kappa$) for particles confined to the $z=0$ plane, there
are many ways to approximate this function by a DF with two
integrals of motion. A frequently used form is
\begin{equation}
    F_0(E,L_z) = \left.\frac{\Omega}{\pi\kappa} \frac{\Sigma_\textrm{d}}{{\tilde\sigma}^2_R}\right|_{R_c} \exp \left( -\frac{E-E_c(L_z)}{\tilde\sigma_R^2(R_c)}\right)\ ,
    \label{eq:dfe}
\end{equation}
where $R_c(L_z)$ and $E_c(L_z)$ are the radius and energy of a
particle on a circular orbit with angular momentum $L_z$. 
$\Omega(R_c(L_z))$ and $\kappa(R_c(L_z))$ are determined by the total gravitational potential, which is a free input function.
The function (Eq. \ref{eq:dfe}) is close to (Eq. \ref{eq:dfv}) when the radial
velocity dispersion $\tilde\sigma_R$ is small, but for real
dispersions the difference in the inner disc can be significant.
Besides, function (Eq. \ref{eq:dfe}) is an even function of $L_z$, thus,
for example, its values on lines of prograde and retrograde circular
orbits coincide. This calls for using various taper functions
curving out a fraction of retrograde and nearly radial orbits, which
add an uncertainty to the study. Another way to account for the
retrograde orbits and to approximate DF (Eq. \ref{eq:dfv}) without using
taper functions is to consider $R_c$ as a function of $E$. Using a
relation $E-E_c \approx (L_c-L)\Omega(R_c)$ valid for small
$\tilde\sigma$, we replace (Eq. \ref{eq:dfe}) with a DF with the form:
\begin{equation}
    F_0(E,L_z) = \left.\frac{\Omega}{\pi\kappa} \frac{\Sigma_\textrm{d}}{\tilde\sigma^2_R}\right|_{R_c} \exp \left( [L-L_c(E)]\frac{\Omega(R_c)}{\tilde\sigma_R^2(R_c)}\right)\ ,
    \label{eq:dfl}
\end{equation}
where now $\Omega = \Omega(R_c(E))$, $\kappa = \kappa(R_c(E))$. 

\subsection{Bar mode of the basic model}

For the global mode calculation we adopt a DF in the form (Eq. \ref{eq:dfl}) that includes both prograde and retrograde orbits; an effective potential corresponding to the averaged radial force, and averaged profile (Eq. \ref{eq:ave_ompr}); the surface density and radial velocity dispersion profiles inferred from the initial distribution averaged over the vertical axis.

Here we use a matrix method by \citet{EP05} that has a form of the
linear matrix equation,
\begin{equation}
    \vA \vx = \omega \vx\ ,
    \label{eq:me}
\end{equation}
which allows us to find unstable modes effectively without a priory
information on the localisation of modes.

\begin{figure}
\centerline{\includegraphics [width = 85mm]{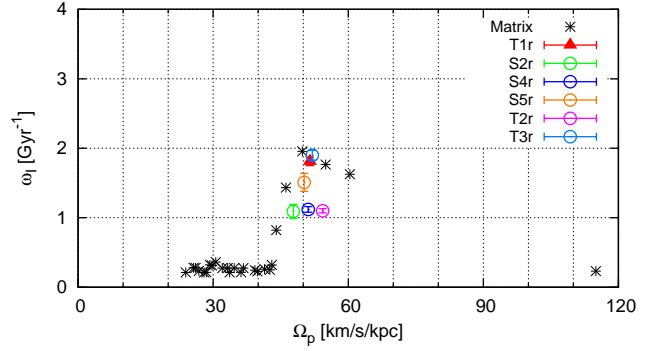}}
\caption{Disc eigenmodes of the basic model obtained with the matrix method.}
\label{fig:spec}
\end{figure}

\begin{table}
\begin{center}
\begin{tabular}{l l l l l l}
\hline
$\Omega_\textrm{p}$ & 60.4 & 55.0 & 49.8 & 46.2 & 44.0  \\ 
$\omega_\textrm{I}$ & 1.63 & 1.77 & 1.95 & 1.43 & 0.82  \\
CR                  & 3.48 & 3.86 & 4.32 & 4.70 & 4.96  \\
OLR                 & 6.37 & 7.00 & 7.71 & 8.28 & 8.67  \\
\hline
\end{tabular}
\end{center}
\vspace{-2mm}
\caption{Unstable global modes (pattern speed $\Omega_\textrm{p}$, and growth rate $\omega_\textrm{I}$ in km/s/kpc $\approx$ Gyr$^{-1}$; resonance radii -- in kpc) for the basic model ($z_\textrm{d}=300$ pc, maximum of the averaged $\Omega_\textrm{pr}$ profile is 43.7 km/s/kpc) obtained with the matrix method.}
\label{tab_basic}
\end{table}

The obtained spectrum consists of two groups of modes (Fig.\,\ref{fig:spec}). The faster
rotating modes with $\Omega_\textrm{p} > 80$ km/s/kpc are localised
between corotation resonance CR and OLR, and they are out of the
scope of what we study here. In the slower rotating group we see
several dominating modes. All unstable matrix modes (Table \ref{tab_basic}) lack the ILR,
which is consistent with the standard point of view on the ILR as a
damping agent. Their growth rates are 0.8 ... 2 Gyr$^{-1}$, i.e.
they coincide with the values obtained in rigid halo/bulge $N$-body simulations. The
$N$-body pattern speed of 52 km/s/kpc is a mean value of the matrix
pattern speeds, so the bar mode can be a superposition of matrix
modes.

\begin{figure*}
\centerline{\includegraphics [width = 170mm]{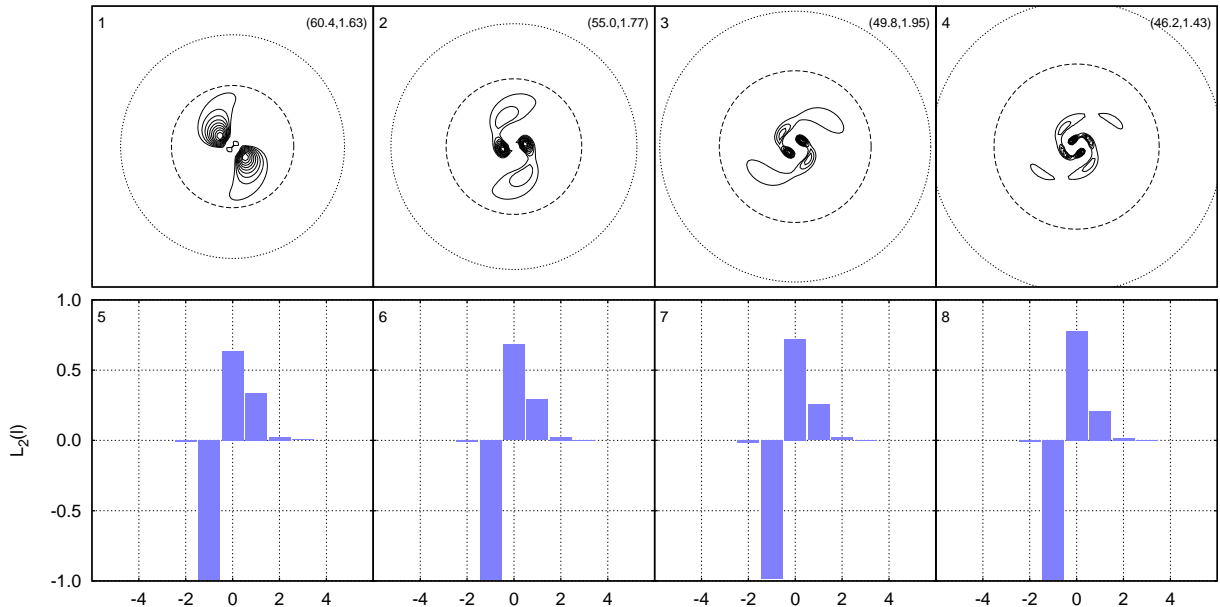}}
\caption{The matrix modes. Frames (1--4): patterns of the most
unstable modes. Isolines show the excess surface density over the
axially symmetric distribution by 10, 20, ... 90 per cent. The
pattern speeds and the growth rates are given in parentheses,
$(\Omega_\textrm{p},\omega_\textrm{I})$. Dashed lines show CR, while
dotted lines show the Lindblad resonances. Each frame size is 16x16
kpc. Bars (5--8) indicate the dependence of $L_2(l)$ vs. $l$
reflecting angular momentum exchange between different parts of the
disc.} \label{fig:pme}
\end{figure*}

Patterns of the matrix modes are presented in Fig.\,\ref{fig:pme}.
They look like lumpy structures with different numbers of
maxima spaced by approximately 90 degrees. Modes with larger growth
rates are more likely to be dominant, thus it is natural to expect
the observed pattern speed to be between 49.8 and 55 km/s/kpc. Each
of the frames shown in Fig.\,\ref{fig:gS2} is a superposition or
nonlinear evolution of these modes.

The bar diagrams below the patterns indicate angular momentum exchange in the modes above. Each bar is proportional to Fourier components of angular momentum $L_m(l)$, normalised so that the sum of the positive components equals one. The Fourier components,
\begin{equation}
L_m(l)  = -\int\d\vJ F'_{0, l}(J) \frac{|\Psi_l (\vJ)|^2}{|\omega - l\Omega_1 (\vJ) - m\Omega_2 (\vJ)|^2}\
\label{am}
\end{equation}
obey the angular momentum conservation law, $L_m = \sum_l L_m(l) =
0$. In the expression, $\Omega_1$ and $\Omega_2$ are radial and
angular frequencies of a particle in the orbit $\vJ = (J_R, L_z)$,
$J_R$ is the radial action, and $L_z$ is the $z$ component of the
angular momentum. Other entities are explained in \citet[][Sect.
2.2]{PJ15}. Contribution to each bar comes mainly from the region
where the corresponding denominator peaks. For $l=-1$ substitution
$\omega = m\Omega_\textrm{p} + \textrm{i}\omega_\textrm{i}$, and
$m=2$ gives the peak where $|\Omega_\textrm{p} -
(\Omega_2-\Omega_1/2)| $ is the smallest, i.e. in the vicinity of
the precession curve maximum. Similarly, $l=0$ and $l=1$ bars
contribute mainly corotation and OLR regions. Thus, we conclude from
the bar charts that the angular momentum exchange occurs mainly
between stars in the vicinity of the precession curve maximum and
stars on corotation and OLR {resulting in a net outward flow of angular momentum}.

\subsection{Different vertical scales}

In this section we compare N-body and matrix modes for the discs
with different thickness controlled by the vertical scale parameter
$z_d$. An initial range of $z_d$ was narrowed to $100 ... 400$ pc.
The disc with $z_d =100$ pc is an outlier in our N-body analysis,
and it suffers from buckling instability. Discs with $z_d > 400$ pc
turned out to be stable.

\begin{figure}
\centerline{\includegraphics [width = 85mm]{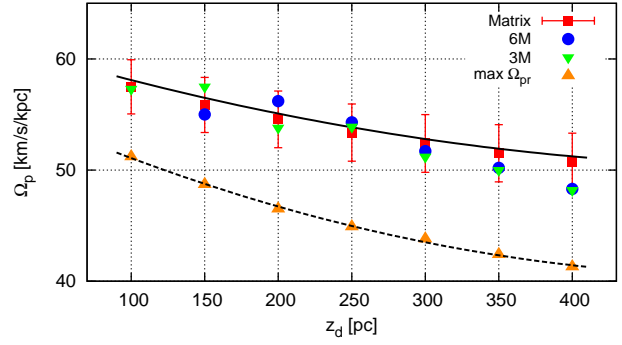}}
\caption{The pattern speeds of matrix and N-body modes calculated in $N_\textrm{d}=6$M and 3M runs. Upper and lower limits of the red error bars show pattern speeds of modes with the largest growth rates. Orange upward triangles show maxima of the precession curves for each $z_d$. Black dashed line shows a smooth fit to the triangles. Black solid curve shows a smooth fit to the average pattern speeds of matrix modes. }
\label{fig_ozd}
\end{figure}

\begin{figure}
\centerline{\includegraphics [width = 85mm]{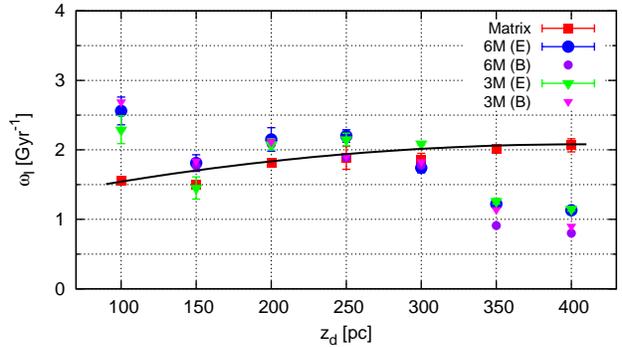}}
\caption{The growth rates of matrix and N-body modes calculated in $N_\textrm{d}=6$M and 3M runs. Upper and lower limits of the red error bars show two maximum growth rates. 6M(E) and 3M(E) show the rates obtained with wave amplitudes (Eqs. \ref{eq:sd1}--\ref{eq:vp2}), while 6M(B) and 3M(B) show the rates obtained with bar strength (Eq. \ref{eq:barstr}) using 6M and 3M N-body runs. }
\label{fig_gzd}
\end{figure}

We performed two series of N-body simulations with axisymmetric halo
and bulge potentials using {\tt ber-gal0} Tree code, with
$N_\textrm{d}=6$M and 3M particles. The obtained N-body modes, along
with the matrix modes, are presented in Tab.\,\ref{tab_zd} and shown
in Figs.\,\ref{fig_ozd}, \ref{fig_gzd}.

\begin{table*}
\begin{center}
\begin{tabular}{l l l l l l l l l}
\hline
Modes          & $z_d$  [pc]                &  100 & 150 & 200 & 250 & 300 & 350 & 400\\ [1mm]
6M             & $\Omega_\textrm{p}$  &  -- & 56.9 & 56.7 & 55.0 & 52.0 & 50.3 & 47.0\\
               & $\omega_\textrm{I}$ (B)&  -- & 2.58 & 2.69 & 2.50 & 1.86 & 1.23 & 1.05\\
               & $\omega_\textrm{I}$ (E)&  $2.56 \pm  0.20$& $1.81 \pm  0.12$& $2.15 \pm  0.17$& $2.20 \pm  0.09$& $1.74 \pm  0.07$& $1.22 \pm  0.01$& $1.13 \pm 0.03$ \\  [1mm]
3M             & $\Omega_\textrm{p}$    &  59.7 & 56.2 & 56.0 & 54.2 & 52.1 & 50.2 & 48.3\\
               & $\omega_\textrm{I}$ (B)&  2.91 & 1.72 & 2.61 & 2.27 & 2.20 & 1.36 & 1.13\\
               & $\omega_\textrm{I}$ (E)&  $2.29\pm0.20$ & $1.45 \pm 0.16$ & $2.08 \pm 0.08$ & $2.15 \pm 0.09$ & $2.09 \pm 0.02$ & $1.26 \pm 0.04$ & $1.15 \pm 0.03$\\  [1mm]
Matrix         & $\Omega_\textrm{p} + \textrm{i}\omega_\textrm{I}$ & 64.8 + 1.25i &  63.2 + 1.43i &  62.3 + 1.54i &  61.2 + 1.55i &  60.4 + 1.63i &  59.5 + 1.56i &  58.9 + 1.65i \\
        & $\Omega_\textrm{p} + \textrm{i}\omega_\textrm{I}$  &59.9 + 1.61i &  58.3 + 1.48i &  57.1 + 1.77i &  56.0 + 1.72i &  55.0 + 1.77i &  54.1 + 1.96i &  53.3 + 1.97i \\
        & $\Omega_\textrm{p} + \textrm{i}\omega_\textrm{I}$  &55.0 + 1.51i &  53.4 + 1.51i &  52.0 + 1.87i &  50.8 + 2.05i &  49.8 + 1.95i &  48.9 + 2.07i &  48.1 + 2.16i \\
        & $\Omega_\textrm{p} + \textrm{i}\omega_\textrm{I}$  &51.4 + 1.08i &  50.0 + 1.30i &  48.4 + 1.44i &  47.2 + 1.32i &  46.2 + 1.43i &  45.3 + 1.30i &  44.5 + 1.45i \\
\hline
\end{tabular}
\end{center}
\vspace{-2mm}
\caption{N-body (6M and 3M) and matrix modes. For N-body modes, growth rates are calculated using both the bar strength (Eq. \ref{eq:barstr}) (denoted by [B]) and wave amplitudes (Eqs. \ref{eq:sd1}--\ref{eq:vp2}), denoted by [E].}
\label{tab_zd}
\end{table*}

Fig.\,\ref{fig_ozd} shows pattern speeds of the modes. For each
$z_d$ we have four modes with the highest growth rates which were
used to calculate a pattern speed as an average weighted by the
growth rates. These pattern speeds are then used to calculate a
smooth fit (black solid curve). Upper and lower limits of the red
error bars show pattern speeds of modes with the largest growth
rates. As expected, matrix modes appear above the maxima of the
precession curves. N-body modes agree well with the matrix
calculations. 

Fig.\,\ref{fig_gzd} shows growth rates of the modes. For each $z_d$
we have shown two maximum growth rates (upper and lower limits of
the error bars). The growth rates were calculated using bar strength
(Eq. \ref{eq:barstr}) and wave amplitudes (Eqs. \ref{eq:sd1}--\ref{eq:vp2}),
in both series of N-body runs. The growth rates obtained with the
two methods are quite close. Comparison with matrix equations shows
good agreement for 150, 250, 300 pc runs. The model with 100 pc is
an outlier, for a reason which is not well understood. In
particular, for 6M run we were unable to determine the pattern speed
and growth rate $\omega_\textrm{I} [E]$. For the thicker discs, $z_d
\geq 350$ pc, N-body shows a decrease of the instability, while the
matrix method shows a weak trend of instability increase. A possible
reason for the different trends is an effect of increasing vertical velocity dispersion that
act similar to the radial velocity dispersion and stabilise the
disc, which is not included in the martix method.

\section{Discussion and Conclusions}

In this paper we study bar formation in a galactic model with a cusp
and other parameters which our Galaxy possibly possesses. The model
consists of three stellar components: exponential disc, cuspy bulge,
and NFW halo. Our study includes $N$-body simulations with
particle--mesh (PM) and Tree codes, and the calculation of global
modes.

First of all, using one `basic' model and more than 100M particles,
we definitely show that a bar really forms despite the presence of
the ILR. This is the case in live and rigid halo and bulge runs.
Maximum bar radius is achieved soon after saturation of the
instability. In the live runs it is around 5 kpc, with  ellipticity
of about 0.75 (Tab.\,\ref{tab_rb_live}). In the rigid runs, the
obtained bars are somewhat shorter and rounder: maximum radius is
4.3 kpc, with ellipticity 0.68 (Tab.\,\ref{tab_rb_rigid}).

Substitution of halo and bulge particles with the rigid external
potential revealed that rigid cuspy models are less unstable than
live ones, i.e. typical growth rates are twice as small. This is in
agreement with results by \citet{A02} obtained for the bulgeless
models. At the same time, pattern speeds in live and rigid runs are
close (relative difference is 5 per cent or less).

Stochastic behaviour of N-body models, mentioned e.g. by
\citet{SD09} for disc evolution after bar formation, is also seen in
our runs, especially in the case of the rigid halo and bulge, when
growth rates are small. This manifests itself in appearance of the
random lag before the exponential growth of the amplitude, and
sometimes in the non-exponential character of the growth. Special
techniques are used to obtain models that demonstrate more evident
exponential growth \citet[e.g.][]{SA86}, but here we avoid using it
emphasising that stochasticity and shot noise are inherent
properties of the stellar systems.

The models demonstrating the exponential growth preserve initial
axisymmetric parameters quite well so that the Toomre $Q$ parameter
remains unchanged almost until saturation. This fact allows one to
calculate unstable global modes using matrix methods of linear
perturbative analysis, and compare them with rigid N-body runs.
While using the cuspy potential in the equatorial plane, no unstable
modes are found by the matrix method \citep{EP05}.

For usual bar mode instability in thin discs, the behaviour of the $\Omega_\textrm{pr}$ profile determining the position of the ILR is important. We emphasise that the resonance radius for the ILR is well defined only for nearly circular motion in razor thin discs. For eccentric orbits in the plane, the ILR is a curve in action space. If the orbit is not planar, frequencies determining the ILR can be approximated if radius $R$ is much larger than the maximum elevation $z$ above the plane. However, if $R \lesssim z_\textrm{d}$, the orbit is essentially 3-dimensional and generally not quasiperiodic. Thus, at radii $R \lesssim z_\textrm{d}$, the resonances are smeared out and the appeal of an ILR is meaningless.

Using a toy model that averages radial forces along the axis
perpendicular to the disc plane, we obtain an effective potential
and $\Omega_\textrm{pr}(R)$ that possesses a maximum. In other
words, the obtained precession curve is typical for cored models,
which allows for the unstable modes. The obtained global modes in
models with various vertical scales in the range $100 \leq z_d \leq
400$ pc agree well with the $N$-body calculations.

Matrix calculations show strong dependence of the growth rates from
the disc mass. For example, for $M_\textrm{d}$ only 11 per cent
lower than adopted in our models, the obtained growth rates are 2.5
times smaller. If one can extrapolate these results to live discs
preserving the ratio of the growth rates of the live and rigid
models at factor of 2, it means that the galactic disc remains
nearly stable (instability time is larger than age of the Universe)
for a long time, and the bar started to form only recently. Our
crude estimates using reasonable star formation rates show that
during 10 Gyr small disc perturbations grow by a factor
$\exp(5...6)$ which is needed for the Poisson noise of the disc
consisting of $N_\textrm{d} \sim 10^6$ particles to grow into a bar.
This fact can explain the observed lack of barred galaxies at
redshifts $z\gtrsim0.5$ \citep{A99, M00}, and typical ratios of
corotation to bar radii, $0.9 < {\cal R} < 1.3$ \citep{BT08}, as a
sign of relatively young bars close to their maximum size.

In this paper we have revealed an ILR is practically non-existent, taking into consideration the thickness of the disc. On the other hand, according to the WASER theory \citep[e.g.,][]{B14}, unstable modes could be excited if there is an inner Q-barrier and such a barrier shields the existing ILR. In fact, a cubic dispersion relation by \citet{B89} gives one `shielded' mode with the pattern speed $\Omega_\textrm{p} = 55$ km/s/kpc, i.e. very close to predictions of N-body simulations and matrix equations (shielding starts from 44 km/s/kpc). In a separate paper we plan to explore WASER mechanism in more detail using models with different radial velocity dispersions.

\section*{Acknowledgments}

The main production runs was done on the {\tt MilkyWay} supercomputer, funded by the Deutsche
Forschungsgemeinschaft (DFG) through the Collaborative Research Centre (SFB 881) ``The Milky Way System'' 
(subproject Z2), hosted and co-funded by the J\"ulich Supercomputing Center
(JSC).

The special GPU accelerated supercomputer {\tt laohu} at the Center
of Information and Computing at National Astronomical Observatories,
Chinese Academy of Sciences, funded by Ministry of Finance of
People's Republic of China under the grant $ZDYZ2008-2$, has been
used for some of code development. We also used a smaller GPU
cluster {\tt kepler}, funded under the grants I/80 041-043 and I/81
396 of the Volkswagen Foundation and grants 823.219-439/30 and /36
of the Ministry of Science, Research and the Arts of
Baden-W\"urttemberg, Germany.

This work was supported by the Sonderforschungsbereich SFB 881 ``The Milky Way System'' (subproject A6) 
of the German Research Foundation (DFG), and by the Volkswagen Foundation under the Trilateral Partnerships 
grant No. 90411. The authors wish to thank G. Bertin, J. Sellwood, I.G.\,Shukhman, J. Wicker for reading the paper and
suggesting improvements to the text, and anonymous referee for valuable remarks and assistance. E.P. acknowledges financial support by the Russian Basic Research Foundation, grants
15-52-12387, 16-02-00649, and by the Basic Research Program OFN-15
`The active processes in galactic and extragalactic objects' of
Department of Physical Sciences of  RAS. P.B. acknowledges the
special support by the NASU under the Main Astronomical Observatory
GRID/GPU {\tt golowood} computing cluster project.

\end{document}